\documentclass[sigconf]{acmart}

\AtBeginDocument{%
  \providecommand\BibTeX{{%
    \normalfont B\kern-0.5em{\scshape i\kern-0.25em b}\kern-0.8em\TeX}}}

\copyrightyear{2023} 
\acmYear{2023} 
\setcopyright{rightsretained} 
\acmConference[CHI EA '23]{Extended Abstracts of the 2023 CHI Conference on Human Factors in Computing Systems}{April 23--28, 2023}{Hamburg, Germany}
\acmBooktitle{Extended Abstracts of the 2023 CHI Conference on Human Factors in Computing Systems (CHI EA '23), April 23--28, 2023, Hamburg, Germany}\acmDOI{10.1145/3544549.3585860}
\acmISBN{978-1-4503-9422-2/23/04}

\usepackage{array}
\usepackage{multirow}
\usepackage{subfigure} 
\usepackage{graphicx}
\usepackage{booktabs}
\usepackage{lipsum}

\begin{document}

\title[Insights on Augmented Reality in Service of Human Operations on the Moon]{Augmented Reality in Service of Human Operations on the Moon: Insights from a Virtual Testbed}

\author{Leonie Becker}
\orcid{0000-0003-4736-5579}
\affiliation{%
  \institution{German Aerospace Center (DLR) - Software for Space Systems and Interactive Visualization}
  \streetaddress{Lilienthalplatz 7}
  \city{Brunswick}
  \country{Germany}
  \postcode{ 38108}
}
\email{Leonie.Becker@dlr.de}

\author{Tommy Nilsson}
\orcid{0000-0002-8568-0062}
\affiliation{%
\institution{European Space Agency (ESA)}
 \streetaddress{Linder Hoehe}
 \city{Cologne}
 \country{Germany}
 \postcode{51447}
}
\email{tommy.nilsson@esa.int}

\author{Paul Topf Aguiar de Medeiros}
\orcid{0000-0003-3785-9778}
\affiliation{%
  \institution{European Space Agency (ESA)}
  \streetaddress{Linder Hoehe}
  \city{Cologne}
  \country{Germany}
  \postcode{51447}
}
\email{hello@pauldemedeiros.nl}

\author{Flavie Rometsch}
\orcid{0000-0001-5827-431X}
\affiliation{%
  \institution{European Space Agency (ESA)}
  \city{Noordwijk}
    \streetaddress{Keplerlaan 1}
  \country{Netherlands}
  \postcode{2201}
}
\email{Flavie.Rometsch@esa.int}

\renewcommand{\shortauthors}{Becker et al.}

\begin{abstract}
Future astronauts living and working on the Moon will face extreme environmental conditions impeding their operational safety and performance. While it has been suggested that Augmented Reality (AR) Head-Up Displays (HUDs) could potentially help mitigate some of these adversities, the applicability of AR in the unique lunar context remains underexplored. To address this limitation, we have produced an accurate representation of the lunar setting in virtual reality (VR) which then formed our testbed for the exploration of prospective operational scenarios with aerospace experts. Herein we present findings based on qualitative reflections made by the first 6 study participants. AR was found instrumental in several use cases, including the support of navigation and risk awareness. Major design challenges were likewise identified, including the importance of redundancy and contextual appropriateness. Drawing on these findings, we conclude by outlining directions for future research aimed at developing AR-based assistive solutions tailored to the lunar setting.  

\end{abstract}


\begin{teaserfigure}
  \includegraphics[width=\textwidth]{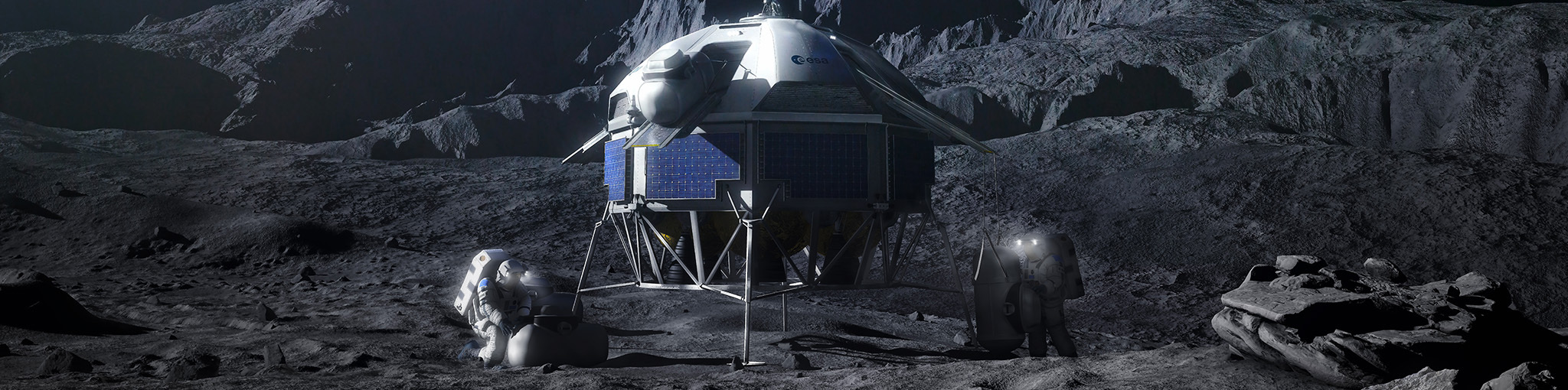}
  \caption{Simulated cargo unloading operation performed in our virtual testbed representing the lunar south pole}
  \Description{Simulated cargo unloading operation performed in our virtual testbed representing the lunar south pole.}
  \label{fig:teaser}
\end{teaserfigure}
\maketitle

\section{Introduction}
The ongoing Artemis missions are setting the stage for expanded human lunar exploration. In collaboration with its international partners, NASA plans to send astronauts back to the Moon in the coming years with the view of establishing a sustainable human presence on its surface by the end of this decade \cite{smith_artemis_2020}. In contrast to past Apollo missions, which primarily involved short-duration activities, such as the collection of geological samples near the lander, future astronauts will need to perform tasks of a more complex nature \cite{haney2020apollo}. 
 These will range from unloading and transportation of cargo supplies to the construction and maintenance of habitation modules \cite{landgraf_lunar_2021}. Accomplishing such undertakings in a safe and effective manner will require the development of dependable assistive technologies capable of underpinning extravehicular activities (EVAs) of future astronauts living and working in the extreme lunar environment.  

This is no small task. As demonstrated by the Apollo missions, future crews will have to face major obstacles impeding their physiological and psychological well-being, work performance and safety. Notably, the lack of atmospheric light scattering on the Moon results in pitch-black shadows and blinding highlights, rendering even seemingly trivial tasks arduous, whilst also elevating safety risks, such as tripping hazards. In addition, the general absence of landmarks and reference points was found to impair distance estimation of astronauts traversing the desolate and achromatic lunar landscape \cite{eppler_lighting_1991, fassett_effective_2020}. Already described as an issue by Apollo astronauts, this factor will be even more pronounced on the lunar south pole - the designated landing site for future lunar expeditions. 

Historically, Apollo astronauts sought to mitigate such challenges by making use of extensive checklists or procedural manuals \cite{hersch2009checklist}, as well as maps featuring complex information to support navigation in the unfamiliar terrain. Such maps were however characterized by displaying the lunar environment from a top-down perspective, prompting the astronauts to rotate the map and search for landmark references in order to navigate, a task that added a significant amount of mental workload to the already challenging conditions \cite{anandapadmanaban_holo-sextant_2018}. To make matters worse, future lunar astronauts will have to deal with limited mobility and field of vision due to wearing EVA space suits, contributing to a lack of situational awareness, high levels of fatigue, and further impairing spatial orientation during EVA operations \cite{davis2019testing}. Additionally, due to communication latencies between Earth and Moon, a high degree of astronaut autonomy will also be required \cite{chappell2017evidence}.

 Against this backdrop, space agencies around the world are currently exploring several assistive technologies in support of future lunar expeditions, ranging from pressurized rovers capable of transporting astronauts between points of interest to a modernized spacesuit that allows for greater mobility \cite{creech2022artemis}. 
 
 Following this vein of inquiry, NASA has likewise identified the use of AR systems as a potential solution to many of the aforementioned challenges \cite{cardenas2021aaron}. By superimposing contextually relevant computer-generated information on the user's view of the real world \cite{coan_exploration_2020}, AR interfaces have in this sense been suggested to potentially help compensate for some of the difficulties associated with EVA operations on the Moon \cite{de_medeiros_categorisation_2022}. Such conjectures are not unfounded. AR technology has already been successfully utilized to support human workflows in numerous domains, such as by enhancing perception and situational awareness of soldiers \cite{livingston_military_2011} \cite{mao2019augmented} or by aiding engineers during complex maintenance and repair tasks \cite{henderson_exploring_2011}. In this regard, the use of HUDs has proven to be particularly advantageous, as the operator is not required to shift his gaze during the completion of tasks since 3D information is projected directly onto the field of view \cite{thomas1992augmented}.

Nevertheless, the potential use of AR in the context of lunar surface operations is still largely unexplored. This absence of empirical research stems predominantly from the difficulties entailed in simulating a representative lunar environment for the purpose of experimental deployments. Whilst several prospective lunar surface technologies have been studied in analog testbeds, such as underwater to simulate reduced gravity \cite{Dorota}, this practice has attracted criticism for being slow, logistically demanding, and oftentimes prohibitively expensive \cite{casini2020lunar}. To address this limitation, we have developed an innovative approach based around digital twinning to recreate relevant lunar regions in VR and subsequently immersively and interactively simulate prospective lunar surface scenarios. This then provided us with the means for making relevant workflows and prospective technologies available for assessment by domain experts.  

In this paper, we present findings from our initial pilot study featuring 6 participants, including experienced astronauts, instructors, and scientists. Drawing on their qualitative reflections, we consider the challenges reported during the Apollo missions and highlight several use cases for AR-based assistive solutions, elaborating on both the opportunities and challenges arising from their use during lunar EVA operations.

The contribution of this work is thus twofold: On the one hand, we demonstrate the methodological viability of employing simulated VR environments as a testbed for the evaluation of prospective lunar surface technologies. On the other hand, we present preliminary findings from an exploratory study on AR and its applicability in the unique lunar context. 

The remainder of this paper is organized as follows: Section 2 examines past research concerned with AR applications for human space flight. The potential use of VR to facilitate the evaluation of design concepts is likewise reflected upon. In section 3 we detail the methodology of our study, followed by an analysis of the findings in section 4. We conclude in section 5 by reflecting on the significance of these findings and propose directions for future research to expand on this pilot study and further explore AR systems in support of future human exploration of the Moon. 

\section{Related Work}
AR’s capacity to visualize complex sets of data and project contextually relevant digital information onto the user’s view of the real world are earning it a growing popularity across fields and disciplines \cite{mohr_retargeting_2015, henderson_exploring_2011}. Workflows centered around AR are now emerging in domains as diverse as physics education \cite{radu2019can}, location-based gaming \cite{paavilainen2017pokemon}, and additive manufacturing \cite{peng2018roma}.  

The utility demonstrated by AR interfaces has not gone unnoticed by the aerospace industry either \cite{safi2019review}. While concepts of  HUDs for astronaut suits were drawn out as early as the 1980s \cite{gernux1989helmet}, experimental deployments of relevant solutions in simulated settings on Earth, as well as on board the International Space Station (ISS), have only begun to take shape relatively recently \cite{byrne2019treadmill, ramsey2015nasa}. These studies have nevertheless already surfaced several promising results. The visual, interactive, and contextual nature of AR, for instance, attracted praise for allowing astronauts to cognitively process information more efficiently than what would be possible via traditional means \cite{mitra2018human}, resulting in lower mental and temporal workload during operations, such as ISS maintenance tasks \cite{braly_augmented_2019, markov-vetter_pilot_2013}.  

Such contextual testing of prototypes is a common practice in relevant user-centered design processes. Whilst the ISS offers an opportunity for experimentation in the low Earth orbit, in the case of space systems for lunar and planetary exploration, analog environments have to be used to approximate the context of use. AR-based solutions for lunar surface operations have thus been tested in various terrestrial settings, including NASA’s Neutral Buoyancy Lab \cite{coan_exploration_2020} and natural analog environments, such as cave systems or other relevant geological sites\cite{lopez-contreras_testing_2022,anandapadmanaban_holo-sextant_2018, rometsch_design_2022}.   

Apart from being costly and logistically demanding, the reliance on such analog testbeds has also attracted criticism for underexposing key elements of the lunar environment, including the unique lighting conditions on the lunar south pole \cite{nilsson_using_2022}. An alternative approach utilizing VR as a testbed to immerse participants in a realistic scenario and elicit context-specific comments and insights from experts has been utilized in other domains \cite{kuliga2015virtual}, but it has not yet been applied to the evaluation of potential applications for AR during lunar surface EVAs. Yet, numerous studies suggest that VR applications can significantly improve the identification of user requirements and needs during an engineering design process \cite{chang2022influence, hubenschmid2022relive, lee2019design, jetter2020vr, nilsson_using_2022}.  

Understanding user needs is key to developing complex systems such as AR displays for EVAs \cite{iso199913407}. Given that significant investments of resources can be expected to be required for the development of an AR system for EVAs, it is important that a good foundation is laid for this work, and the potential use cases, challenges, and requirements are explored before the system is designed. Drawing on literature analysis and brainstorming sessions, for instance, De Medeiros et al., \cite{de_medeiros_categorisation_2022} formulated a categorization of applications for AR in human lunar exploration. This included categories such as 'EVA navigation', 'sample collection', 'maintenance, repair, overhaul and construction', 'medical procedures', and 'biomedical and system status monitoring'. Such categorization can be used to relate individual findings from separate studies which might focus on or be biased towards specific scenarios or use cases to a wider framework in order to gain a broad insight into the potential value of AR systems for lunar EVAs. Our study should therefore reflect on the findings of \cite{de_medeiros_categorisation_2022}, supporting existing categories with more detailed insights on functionalities and design considerations. Or, if functionalities found in this study are not covered by an existing category, suggest  an extension of or alternative to the proposed categorization.

\section{Methodology}
In order to shed light on the potential role of AR in future human operations on the Moon, we have produced a representative virtual testbed and evaluated it with a group of expert users. Below we elaborate on this approach in detail. 
\subsection{The Virtual Testbed}
Drawing on pre-existing topographic scans of the Moon \cite{smith2017summary}, we recreated virtually an area of 64 km\textsuperscript{2} in close vicinity of the Shackleton crater on the Lunar south pole (89.9°S 0.0°E). We selected this area due to it being one of the candidate landing sites for the first Artemis human landing mission \cite{Smith2020}, thus providing a sufficiently plausible backdrop for our study. The Sun was placed at an angle of 1.5° above the horizon to mimic the conditions on the lunar south pole \cite{vanoutryve2010analysis}. All forms of indirect lighting and light scattering were disabled to recreate the pitch-black shadows stemming from the lack of lunar atmosphere.  

While closely following a concept of lunar operations outlined by researchers from ESA and NASA \cite{landgraf_lunar_2021}, we then used this virtual moonscape as a basis for the simulation of a routine cargo reception operation. Such logistical procedures are expected to form a recurring event in future lunar endeavors and therefore constitute a representative operational scenario \cite{sherwood2019principles}. A centerpiece of the virtual environment was a bespoke 3D model of the European Large Logistics Lander (EL3) - an autonomous vehicle for delivery of supplies and other cargo in support of future lunar ground crews, currently under development by ESA \cite{landgraf2022autonomous}. The virtual EL3 was carrying a set of 4 cargo containers that could be accessed by climbing up on top of the lander using an attached ladder. 

To further strengthen the authenticity of the virtual experience, the user was embodied in an accurate virtual representation of an xEMU spacesuit featuring a helmet that partially restricted their field of view \cite{ross2018nasa}. The suit likewise featured two helmet lights. 

The virtual testbed was instantiated using the Unreal Engine 4 game engine and ran on a desktop computer equipped with a RTX 2080 GPU. A HTC Vive Pro VR headset along with two HTC Vive base stations and controllers were employed to facilitate users' interaction with the VR environment. The controller trackpads were used for walking and turning, while the controller triggers were used for interacting with objects (e.g. grabbing and manipulating cargo containers). 
\begin{figure}[h]
\subfigure[]{\includegraphics[width=0.46\textwidth]{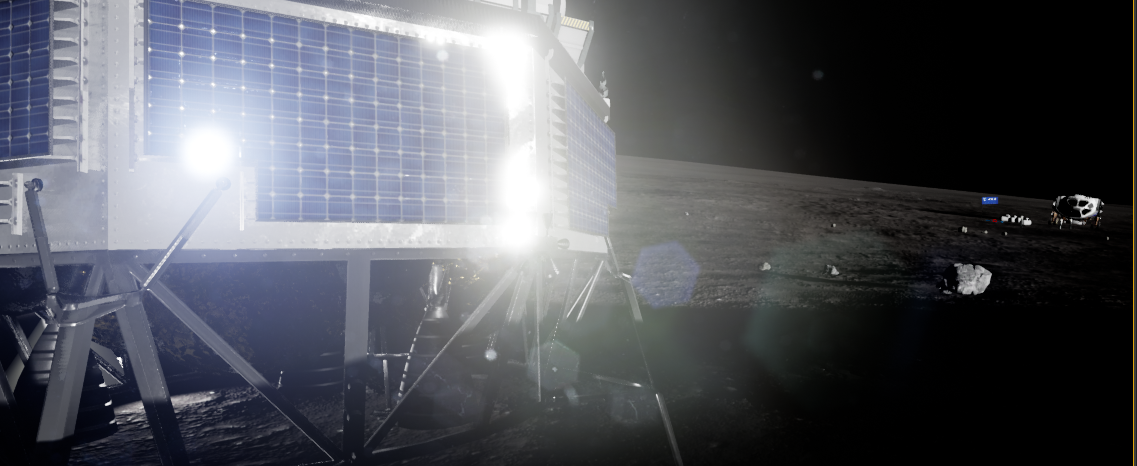}} 
\subfigure[]{\includegraphics[width=0.47\textwidth]{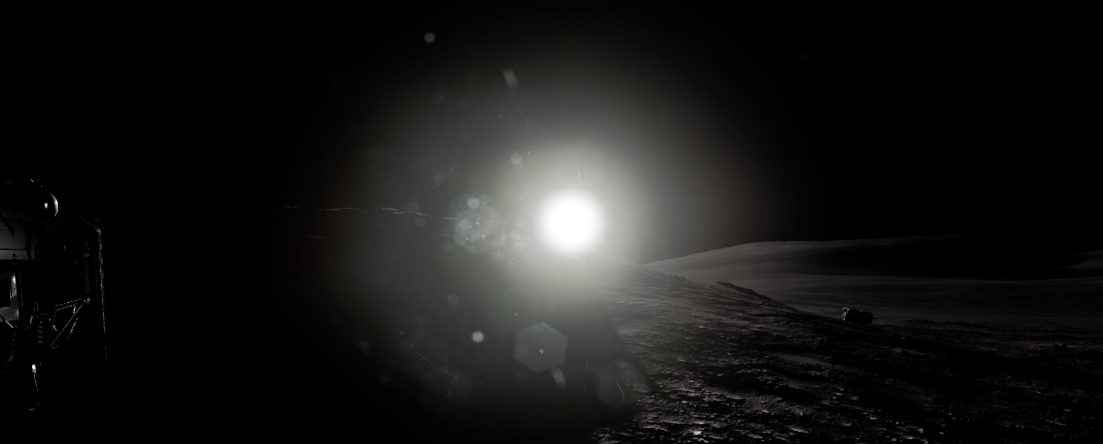}}
  \caption{Virtual lunar surface environment demonstrating strong light reflections (a) and black shadows (b).}
  \Description{Both images show a screenshot of our virtual testbed demonstrating the difficult lighting conditions on the lunar south pole, displaying extreme dark areas and areas showing strong lighting reflections that could blind  astronauts during future lunar missions}
\end{figure}

 \subsection{Participants}
The highly specialized focus of the study prompted us to hand-pick and extend invitations to relevant experts in the aerospace field. We sought to assemble a group with expertise in EVA operations and the pertinent technologies.  

\begin{table*}[hbt!]
    \centering
    \begin{tabular}{|c|c|c|c|}
    \hline
        Participant &  Gender & Job Title & Area of Expertise \\
        \hline

        A1 & M & Astronaut & EVAs / Space Flight \\
   
        A2 & M & Astronaut & EVAs / Space Flight \\
       
          AT1 & F & Astronaut Trainer & EVA Training\\
           
           AT2 & M & Astronaut Trainer & Astronaut Training \\
           
            G1 & M & Geologist & Planetary Exploration / Field Studies / Analog Missions \\
             
             E1 & M & Aerospace Engineer & Field Studies / Analog Missions \\
         \hline    
    \end{tabular}
    \caption{Overview of study participants}
    \label{tab:particpants}
    \Description{Table describing the characteristics of our domain expert participants that were involved in the user study}
\end{table*}

Six participants completed the study (5 male,1 female; see table \ref{tab:particpants}.). Two active astronauts took part in the study. Each of them has recorded approximately 1 year in space on board the ISS across several missions. Astronaut 1 (A1) completed over 33 hours of EVA operations, whereas astronaut 2 (A2) totals around 6 hours.  Astronaut trainer 1 (AT1) is an instructor with over 10 years of experience preparing astronauts for their missions to the ISS whilst also supporting operational monitoring at the EUROCOM flight control center. Astronaut trainer 2 (AT2) is an expert in astronaut EVA training with first-hand experience in carrying out simulated astronaut operations in different types of spacesuits. Geologist 1 (G1) holds a PhD in astrophysics with a focus on planetary and lunar geology. He has 15 years of experience working with the world’s largest telescope (GranTeCan) and frequently partakes in simulated analog geological studies. Engineer 1 (E1) is a research scientist with experience of organizing multiple analog field studies for astronaut training.

\subsection{Procedure}

Participants were invited to complete the VR study session individually. Each session started out with the participant being placed within a walking distance of the EL3 mockup. We explained that the participant’s mission is to approach the lander and retrieve one of the cargo containers it was carrying. Next, the participant was instructed to transport the cargo container to a nearby drop-off point, marked by a flag. 

Drawing on the think-aloud protocol \cite{ericsson_verbal_1980}, we encouraged each participant to verbalize their reasoning throughout this procedure. Once they completed their mission, each participant was likewise prompted to answer a set of semi-structured interview questions concerning key aspects of their virtual experience. The aim of these questions was to assess any perceived safety or practical concerns and potential application areas for AR technology on the lunar surface. 

The duration of the sessions varied, ranging from 30 to 45 minutes. All sessions were recorded on audio. Notes were likewise taken throughout the study. A qualitative thematic analysis \cite{braun_using_2006} was then conducted to identify key themes pertaining to potential use cases of AR applications for future human lunar exploration.

\section{Preliminary Findings}

The VR-based testbed was seen as accessible and intuitive, with all of our participants being able to successfully complete their mission. Much like Apollo astronauts before them, all of our participants did however also experience difficulties when confronted with the (simulated) lunar conditions. When prompted to reflect on the potential use of HUD-based AR to mitigate these difficulties, participants were able to identify several qualities that would indeed make AR suitable for aiding future lunar ground crews. A1, for instance, drew on his experience from piloting military jets, and could thus relate to the idea of relying on HUDs for a range of tasks. Similarly, G1 praised the visual nature of AR, arguing this would allow AR interfaces to complement astronaut’s voice communication with the mission control center without interfering with it. 
Overall, participants agreed that AR interfaces could meaningfully enhance astronaut safety and work efficiency during future EVAs. Nevertheless, as participants began to reason about the design and implementation of such interfaces, a more nuanced picture started to emerge, with the need for contextual appropriateness and redundancy being frequently brought up as a requirement. In the remainder of this section, we elaborate on these challenges and opportunities in greater detail. 

\subsection{Astronaut Safety}

\subsubsection{Navigation}
A difficulty of navigating through the lunar landscape was the most immediate obstacle faced by our participants. The unique lighting conditions, with large areas covered in pitch-black shadow, along with the lack of atmosphere and the resulting impairment of distance estimation, made navigation a perceived safety hazard. As some participants suggested, by underestimating distances, astronauts could, for instance, end up wandering too far off from their base, potentially risking running out of oxygen or other vital resources. 

Against this backdrop, all participants agreed that AR-based navigation support could prove crucial. G1 argued an AR interface, drawing on a relevant database of lunar surface maps, should be employed to provide astronauts with directions, helping them find the safest path towards their destination. As G1 elaborated, even a simplistic AR visualization, such as a directional arrow, azimuth information, or a highlighted point of interest, could in this sense significantly improve astronaut’s spatial orientation. 
Similarly, AT2 and G1 both argued that an AR interface, coupled with a LiDAR system, could effectively visualize the distance to surrounding landscape objects, thus providing further support during wayfinding. A2 went a step further, suggesting that to support distance estimation, AR could project 3D models of mundane objects, such as trees and houses, into the real environment: 
\textit{“Because that's how we judge distances down here (on Earth), right? If you are in the Alps and you look at the neighboring mountain, the way you can tell how far away it is, is if you see the houses are really small there. Right?”}. 
AR was in this sense seen as a potential way of bridging the gap between terrestrial settings and the lunar environment by simulating some of the environmental cues we would normally rely on. 

\subsubsection{Collision Prevention}
Apart from aiding navigation, AR was likewise seen as a suitable interface for alerting astronauts to more imminent threats. A2, for instance, suggested that astronauts operating in poor lighting conditions would benefit from nearby rocks, and other sharp objects, being highlighted in order to prevent collisions that could otherwise damage the spacesuit. Similarly, AT1 argued astronauts should be made aware of any moving equipment, such as cargo containers being unloaded, and prompt the astronauts to maintain safe distance. 

AR was nevertheless not always accepted as a self-evident remedy to such safety hazards. E1 expressed doubt concerning the efficacy of an AR interface in collision prevention, arguing that simply providing the astronauts with stronger helmet lights would be a more effective way of increasing their safety. Similarly, AT2 suggested providing all moving equipment with reflective strips or signal lights, in combination with continuous monitoring by the mission control center, which would be adequate to safeguard the astronauts. 
\subsection{Work Performance}

\subsubsection{Work Procedures}
Another major use-case for AR that surfaced through our study was the support of astronaut’s work procedures. 
A2, for instance, argued AR is well suited for facilitating communication and providing real-time instructions during repair operations: \textit{“So I'm here and you're telling me ‘hey, there's a bolt on one of those winches that needs to be tied up here’. And I’m like ‘which one?’. But now my HUD just marks that bolt with a red circle here. And I just see ‘Okay, this one. Yep.’ Or to figure out if it’s the left or the right bolt… Things like that would be super helpful.”}. 
In addition, he argued a HUD providing relevant instructions or checklists could potentially be employed as a step-by-step walk-through guiding astronauts through specific workflows: 
\textit{“Or what do I need to turn my PGT… my electric screwdriver on, what torque settings do I need to turn it up to? So that I don't have to ask the ground (mission control center). I’d just read it from that little list of the procedures that's also in my HUD, things like that. It would be fantastic.”}.
AR interfaces could in this sense be used to lower the dependency of lunar ground crews on continuous instructions provided by the mission control center on Earth, thus increasing their autonomy. Given the risk of communication delays and other interference, such a solution might prove vital for future expeditions.

\subsubsection{Geological Surveys}

In addition to aiding repair tasks, G1 and E1 also both felt AR technology could help astronauts identify points of interest during geological surveys. AR, they explained, is uniquely predisposed to assist astronauts in bulky spacesuits while they are trying to locate and identify strategically important minerals that are not globally distributed, such as titanium or ilmenite, both crucial for future in-situ resource utilization on the Moon. 
A suit-mounted infrared light, G1 explained, could be employed to illuminate nearby rocks, which would in turn enable a spectral analysis and identification of relevant minerals. The location of such minerals could then be communicated to astronauts via AR markers superimposed on the terrain, thus reducing the time and effort they would otherwise have to expend during such procedures. 

\subsection{Key Design Challenges}

\subsubsection{Contextual Relevancy}
As the test sessions matured, it gradually became evident that the wide-ranging applicability of AR identified by our participants constitutes a double-edged sword. As acknowledged by all participants, whilst oftentimes useful, high reliance on AR elements would increase the risk of cluttering and potentially even overwhelming the user’s view, thus largely defeating the purpose of its use. G1 summarized this problem, saying, \textit{“don't make it an Internet Explorer, with multiple tabs playing at the same time”}. Similarly, AT1 stressed the importance of AR interfaces providing users with information that is both contextually relevant, and that cannot be plainly inferred by the user through other means: \textit{”Highlighting that there is a rock in front of you when you can already see the rock, that does not really add anything”}.  

A major design challenge thus revolves around designing AR systems that walk the oftentimes fine line between being informative without being unsolicited. As A1 put it, \textit{“I would not want to saturate my view (…), it must be a good compromise. That is something that needs to be researched. What is the information that’s necessary, and which information is not necessary?”}. A workaround frequently proposed by our participants revolved around maximizing the customizability of AR interfaces by granting the user an option to easily toggle individual features on and off. Not all participants, however, agreed. As explained by A2, manually configuring an interface while wearing a bulky spacesuit would add substantial workload. AT2 likewise stressed that some critical safety-related features, such as warnings concerning imminent collisions, should be made impossible to override.

In order to keep the amount of displayed AR content manageable, several participants suggested additional wearable devices, such as tablets, ought to be utilized to display non-essential information. Indeed, the need to design any AR interface with the broader technological ecosystem in mind surfaced as a recurring theme. For instance, as AT1 elaborated, \textit{“the technical realization of an AR system heavily depends on the development of the astronaut suit that will be used during future lunar missions”}. 

\subsubsection{Technical Redundancy}

Of equal importance, however, is to preserve system functionality, should any of its components experience malfunction. The need for an AR solution to be designed in conjunction with other relevant technologies, whilst maintaining redundancy by avoiding high levels of interdependence, thus emerged as a second major design challenge. As A1 summarized: \textit{“if we can do it and if it can be done in a way that is reliable and if we can have a backup system, then AR could be very positive in my opinion.” }

\section{Discussion}
Thanks to their capacity to vividly superimpose relevant visual information on users' view of real environments, thus facilitating the display of instructions, warnings, and other information in an interactive and time-efficient manner, AR applications are increasingly valued in the space domain. Here, the use of HUDs that do not require a shift of gaze during task completion has been proven particularly useful. Against the backdrop of unique conditions on the lunar surface, such as challenging lighting (e.g., pitch-black shadows) and other ergonomic constraints (e.g., limited field of view and high workload during EVAs), our team explored the potential usefulness of AR technology in this context via the input from 6 aerospace experts. This topic was previously underexplored due to the difficulty in simulating the lunar environment and the specific lighting conditions on the Moon's surface. To address this gap in research, we created a VR simulation of a realistic lunar environment encapsulating some of the unique lunar conditions. 

The results of our study indicate that supportive AR HUD technology could address some of the problems faced by the Apollo crews, such as challenging lighting conditions and difficulties to judge size and distance of objects and landmarks by spatially displaying navigational cues on the lunar environment. Also, the domain experts indicated that hazard warnings displayed in the shadow could increase safety of future lunar astronauts. Moreover, the use of supportive AR HUD systems to display instructions emerged as a topic in the discussion with the participants, showing that AR could provide astronauts with more user-friendly instructions than during previous missions where extensive (cuff-) checklists had to be used. Our findings also indicate that an AR system might allow for interactive collaboration between mission control center and the astronauts. Supportive AR technology could be furthermore utilized to display relevant data about the astronaut suit, a topic that was, for instance, also explored by \cite{mitra2018human}.  

These use cases identified by our study are largely congruent with the classification of potential AR applications described by De Medeiros et al. \cite{de_medeiros_categorisation_2022}. Functions related to \textit{navigation} and \textit{collision prevention}, for instance, correspond to the ‘EVA Navigation’ category, while the \textit{work procedures} use-case corresponds to the ‘Maintenance, repair, overhaul and construction’ and the 'Collaboration and support' categories.

Nevertheless, the detailed insights provided by participants in our study can add to such body of work. Although the somewhat generalized categories from \cite{de_medeiros_categorisation_2022} can indeed prove useful in enabling a high-level classification of findings made by various enquiries into lunar AR, they cannot replace the qualitative, highly specific and context-dependent design challenges and functionalities which were suggested by participants in this study. For instance, the risk of overwhelming users with unnecessary information and the need to adopt systems thinking whilst designing for redundancy attracted considerable attention in our study. Such findings extend the research of \cite{de_medeiros_categorisation_2022} by demonstrating the importance of a contextual deployment and evaluation to reveal underlying complexities, such as the interoperability between multi-device ecosystems and potential implementation issues. 

In addition to exploring potential applications and design considerations for AR technologies in the context of future lunar EVA missions, as a secondary  contribution, our work thus demonstrates the viability of virtual testbeds as a methodological tool for facilitation of contextual enquiries concerning prospective and hypothetical systems.

Below we discuss and reflect on potential areas for future work that could extend our findings: 

\begin{itemize}

\item\textbf{ Extension of AR Categories for Lunar Exploration} -  Future studies should further explore and extend the categories proposed by \cite{de_medeiros_categorisation_2022}. Here, our findings especially highlight the importance of user studies involving domain experts as participants to investigate potential AR applications for lunar EVAs. In future studies, we thus plan to systematically explore and reflect on the categories proposed by \cite{de_medeiros_categorisation_2022} utilizing our virtual testbed. 
\\

\item \textbf{Integration of AR Features Into the Virtual Testbed} 
-  One limitation of our study was that participants had to imagine how AR technology implemented into a HUD system could assist them during future lunar EVAs. Therefore, as a next step, we are currently implementing several of the suggested AR applications in our virtual testbed environment to further explore the design and usefulness with the participation of additional domain experts. In addition, we intend to investigate existing AR solutions based on current research and plan to explore how these findings can be incorporated into the domain experts' ideas and suggestions.
\\

\item \textbf{Contextual Relevance} - Our work revealed that additional research is required to ensure that only pertinent and context-appropriate information is displayed.  We plan to use eye-tracking technology to determine whether it could serve as a viable interface between augmented reality features and astronauts. Eye-tracking data could serve as the basis for displaying information based on the visual attention of the astronaut. Moreover, it could additionally be used to adjust the appearance of AR applications based on the astronaut's current workload and/or situational awareness \cite{becker2022electroencephalography}.
\\

\item \textbf{Technical Capabilities} - The technical realization and integration of supportive AR systems is another topic that frequently emerged as a point of discussion. Here, most prior research utilized already existing standalone technologies that are not integrated into the astronaut suit (e.g., Microsoft HoloLens) and do not analyze, compute and display data “online” in real-time (e.g., \cite{anandapadmanaban_holo-sextant_2018,rometsch_design_2022}). Future inquiries should thus seek to develop and integrate systems that are capable of displaying various AR features and enable their assessment in a representative technological and environmental context (real or virtual). In this regard, for instance, \cite{mitra2018human} developed a HUD prototype that could show information about the astronaut suit (e.g. oxygen- or battery status), yet the system was not able to display more advanced spatial features, such as navigational cues. Future research should explore how various sub-systems required for a specific AR solution (e.g., detection of hazards or landmarks) can be technically integrated into such a system and how reliable interoperability between its components can be ensured during future missions to the Moon and beyond. 
\\

\end{itemize}

\section{Conclusion}

This pilot study has explored potential AR HUD use cases in support of future human exploration of the Moon by simulating a plausible lunar operational scenario in a virtual testbed. Our findings indicate that AR technology could indeed help mitigate some of the adversities faced by future astronauts, such as extreme environmental conditions impeding their operational safety and performance (e.g. challenging lighting situation or high workload). Domain experts indicated that especially the use cases of navigational cues and the use of AR to display work-related instructions, as well as hazard warnings, could help safeguard future astronauts. Concerns were raised, however, regarding the contextual appropriateness of displayed data, the risk of information overload, and the need for redundancy. To maximize usability and safety during future missions to the Moon and beyond, future research should therefore further explore and expand on these findings. 

\begin{acks} 
 We would like to thank Andreas Gerndt (DLR) and Aidan Cowley (ESA) for their continuous support! In addition, we would like to express our gratitude to the domain experts and astronauts at the European Astronaut Center (EAC) for allowing us to get a deeper understanding of the many requirements and challenges surrounding future lunar EVAs. 
\end{acks}
\balance

\bibliographystyle{ACM-Reference-Format}
\bibliography{bibfile}

\end{document}